\documentclass[12pt,preprint]{aastex}

\newcommand{\solm}{M$_{\odot}$}
\newcommand{\kms}{km\,s$^{-1}$}

\slugcomment{}

\shorttitle{Black Hole in IRS\,13E?}
\shortauthors{Sch\"odel et al.}

\begin{document}


\title{A Black Hole in the Galactic Center Complex IRS~13E?}


\author{R. Sch\"odel, A. Eckart, and C. Iserlohe}
\affil{I.Physikalisches Institut, Universit\"at zu K\"oln, Z\"ulpicher
Str.77, 50937 K\"oln, Germany}
\email{rainer@ph1.uni-koeln.de, eckart@ph1.uni-koeln.de, iserlohe@ph1.uni-koeln.de}

\and

\author{R. Genzel\altaffilmark{1} and T. Ott}
\affil{Max-Planck-Institut f\"ur extraterrestrische Physik,
  Giessenbachstra{\ss}e, 85748 Garching, Germany}
\email{genzel@mpe.mpg.de, ott@mpe.mpg.de}

\altaffiltext{1}{Also: Department of Physics, University of
  California, Berkeley, CA 94720}

\begin{abstract}
The IRS~13E complex is an unusual concentration of massive, early-type
stars at a projected distance of $\sim$0.13\,pc from the Milky Way's
central supermassive black hole Sagittarius~A* (Sgr~A*). Because of
their similar proper motion and their common nature as massive, young
stars it has recently been suggested that IRS~13E may be the remnant
of a massive stellar cluster containing an intermediate-mass black
hole (IMBH) that binds its members gravitationally in the tidal field
of Sgr~A*. Here, we present an analysis of the proper motions in the
IRS~13E environment that combines the currently best available data
with a time line of 10~years\footnote{Based on observations at the
Very Large Telescope (VLT) of the European Southern Observatory (ESO)
on Paranal in Chile}. We find that an IMBH in IRS~13E must have a
minimum mass of $\sim$10$^{4}$\,\solm in order to bind the source
complex gravitationally. This high mass limit in combination with the
absence so far of compelling evidence for a non-thermal radio and
X-ray source in IRS~13E make it appear unlikely that an IMBH exists in
IRS~13E that is sufficiently massive to bind the system gravitationally.
\end{abstract}


\section{Introduction}


In spite of factors such as a strong
tidal field due to the central supermassive black hole Sgr~A*
\citep[e.g.,][]{Schoedel2002Natur,Ghez2003ApJ}, or strong
stellar winds, which pose serious obstacles for star formation,
surprisingly, numerous young (a few times $10^{6}$\,yrs), massive
stars can be found in the central parsec, e.g., the one to two dozen
bright so-called He-stars \citep[with characteristic He\,I emission
lines, see][]{Krabbe1995ApJ,Paumard2001A&A}. The young, massive stars
are mainly concentrated in the region $\sim$10$''$ in projection
around Sgr~A* and appear to rotate around Sgr~A* in two
counter-rotating disks that contain an apparently coeval population of
Wolf-Rayet (WR), Luminous Blue Variables (LBV), and O/B-stars
\citep[][also Paumard et al. 2005, in
preparation]{Genzel2003ApJ,Levin2003ApJ}.
Also, there are B-type main sequence stars in the immediate
vicinity of Sgr~A* \citep[e.g.,][]{Ghez2003ApJ,Eisenhauer2005ApJ}.
The presence of the young, massive stars near Sgr~A* is not well
understood and various explanations for their existence are currently
being discussed, such as, e.g., infall and collision of molecular
clouds, infall and dissolution of a massive cluster \citep[e.g., see
discussions in][]{Genzel2003ApJ}, or star formation in a
self-gravitating accretion disk
\citep[e.g.,][]{Nayakshin2004,MiloLoeb2004ApJ}.

In this context, the IRS~13E complex is a very intriguing
object. Located $\sim3.5''$ in projection from Sgr~A*, it comprises
almost half a dozen massive stars within a projected radius of
$\sim$0.25$''$. These stars appear to be WR stars or O-type
supergiants
\citep[e.g.,][]{Krabbe1995ApJ,Eckart2004ApJ,Maillard2004A&A}. In
addition to representing an unusual concentration of bright,
early-type stars, the IRS~13E sources also have very similar proper
motions \citep[see, e.g., Fig.~18 in][]{Genzel2003ApJ}. 
\citet{Maillard2004A&A} suggested that the IRS~13E complex is the remnant
core of a massive star cluster that had fallen into the central parsec
and dissolved there. Their main arguments are a) the similar proper
motions of the IRS~13E cluster members and b) their finding that the
stars in IRS~13E appear to be massive, short-lived, and therefore
young. 
Since one would expect that an association of stars such as IRS~13E
should be disrupted by the tidal field of Sgr~A*, they speculated that
an intermediate-mass black hole (IMBH) in IRS~13E may prevent its
disruption.  An IMBH in the core of an infalling cluster would also
provide an effective means of inward transportation through increased
dynamical friction as has been suggested by \citet{Hansen2003ApJ}
\citep[see, however,][]{Kim2004ApJL}.

The possible existence of an IMBH in the GC is currently a hypothesis
of high interest in the field. Therefore we consider it timely to
present the newest data on stellar dynamics in IRS~13E.
In this letter, we analyze proper motions of the stars in and near the
IRS~13E complex, derived from ten years of near-infrared (NIR) speckle
and AO observations of the GC stellar cluster.

\section{Observations and Data Reduction}

Observations of the GC stellar cluster are routinely performed since
spring 2002 with the CONICA/NAOS NIR camera/adaptive optics (AO)
system at the ESO VLT unit telescope~4 on Cerro Paranal in Chile.  For
this work, we used K-band (and some H-band) imaging data. After
standard data reduction (sky subtraction, dead pixel correction, flat
fielding, shift-and-add) the final images were Lucy-Richardson
(LR) deconvolved and beam restored with a Gaussian beam of $\sim$60\,mas
(40\,mas for H-band images, respectively), corresponding to the
diffraction limit of the VLT at $2.2\,\mu$m. Stellar positions were
extracted with \emph{StarFinder} \citep{Diolaiti2000A&AS}. They were
transformed into a coordinate system relative to Sgr~A* with the aid
of 18~reference stars with well known positions and proper motions,
taken from \citep{Ott2004PhDT}. Each of the data sets was divided into
two parts that were reduced and analyzed independently. Uncertainties
of the source positions were thus obtained by a comparison of two
independent measurements for each epoch.

We added to these data Gemini North AO H and K imaging data from July
2000. The images are part of the Gemini North Galactic Center
Demonstration Science Data Set. Both images were LR deconvolved and
beam restored. Source positions were extracted in the way described
above (using the two independent measurements provided by the H and K
images).  In order to obtain a larger time baseline, we also used
SHARP speckle imaging data from July 1995, June 1996, and June
2001. Details on the observation and reduction of SHARP imaging data
may be found, e.g., in \citet{Eckart1997MNRAS}. Again, we applied LR
deconvolution and used two independent data sets for each epoch.

\section{Dynamics in the IRS~13E complex}

The IRS~13E sources are named in Fig.\,\ref{Fig:Propmot}.
\citet{Eckart2004ApJ} label E3 and E4 as E3c and E3N,
respectively. \citet{Maillard2004A&A} noted that E3 is a multiple
source and named the two components E3A and E3B.  On the image
presented here, E3 appears to have more than two components. We just
label the brightest one of them E3 (the image in
Fig.~\ref{Fig:Propmot} is based on data with a Strehl  $\geq40\%$
in contrast to the Gemini image of Strehl $\sim$5\% used by
\citet{Maillard2004A&A}).  As concerns E5, it appears highly confused
with other sources and/or extended.

Proper motions of the stars were determined by linear fits to the
measured time-dependent positions with their respective uncertainties
\citep[adopting a GC distance of 7.9~kpc from ][]{Eisenhauer2003ApJ}.
In Fig.~\ref{Fig:Propmot}, the derived proper motions of all stars in
and near IRS~13E are shown superposed on an image. The common proper
motion of the IRS~13E cluster members stands out clearly.  We identify
five stars within a region of about $0.5''$ diameter that could be
part of IRS~13E (see also Table~\ref{Tab:Vels}). There may be further
potential members, but they are either too weak, embedded in extended
emission, or too close to brighter stars, such that no reliable proper
motions could be determined.  The proper motion of E6, while pointing
into the same direction as the ones of the other IRS~13E sources, is
significantly smaller in magnitude. Also, \citet{Maillard2004A&A} note
that E6 is extincted much less than the other IRS~13E stars. It is
therefore questionable whether E6 forms part of IRS~13E and we omit it
from the following analysis.  The proper motion of E5 could not be
reliably determined.

\citet{Maillard2004A&A} considered that IRS~13E might be bound by an
IMBH. From spectroscopically estimated radial velocities for E2 and E4
and assuming that the hypothetical IMBH is located halfway between the two
stars, they estimated its mass to $\sim$1000\,M$_{\odot}$. With the
proper motions, we can derive the minimum mass that is required to
bind the IRS~13E cluster members.  Here, it is first necessary to
subtract the systemic motion of IRS~13E. Subsequently, the residual
velocities of the individual sources can be used for estimating the
mass in IRS~13E under the assumption that it is a gravitationally
bound system. Since we cannot know the true systemic velocity, we
examined three cases: a) Using the average proper motion of the
ensemble, b) using the proper motion of E1, and c) using the proper
motion of E4. Thus, all realistic cases, even extreme ones, should be
covered. In the following, we limit our analysis to case a), which is
the most conservative one because it results in the lowest enclosed
masses.

The average proper motion of IRS~13E, derived from the mean of the
velocities of sources E1, E2, E3, and E4, is $248\pm25$\,km\,s$^{-1}$
westwards and $80\pm48$\,km\,s$^{-1}$ northwards. Here, we chose an
unweighted average because we cannot know which one of the individual
proper motions corresponds best to the possible systemic velocity.
The proper motions of the IRS~13E sources after subtraction of a
systemic velocity corresponding to their mean motion are shown in the
left panel of Fig.~\ref{Fig:Masses}.

A first rough estimate of the mass needed to bind IRS 13E can be
obtained from the residual velocities, through $M=\langle v^2\rangle
R/G$, where $G$ is the gravitational constant and $R\approx10$\,mpc the
size of the system.  With $\langle v^{2}_{RA}\rangle \approx50$\,\kms
($\langle v^{2}_{Dec}\rangle \approx100$\,\kms) we obtain
$M\approx5600$\,\solm ($M\approx22500$\,\solm).

We chose two more refined approaches to estimating the mass of a
putative IMBH that gravitationally binds IRS~13E: a) Estimating the
mass in IRS~13E with the Leonard-Merritt (LM) mass estimator
\citep[eq.~19 in][]{LeonardMerritt1989} and b) deriving a lower limit
on the mass through the relation
$\frac{Rv_{\mathrm{proj}}^2}{2G}\leq\frac{rv_{\mathrm{esc}}^2}{2G}=
M$, with $R$ being the projected distance of a star from the IMBH,
$v_{\mathrm{proj}}$ its velocity projected onto the plane of the sky,
$r$ its real distance from the IMBH, $v_{\mathrm{esc}}$ its escape
velocity, $G$ the gravitational constant, and $M$ the mass of the
IMBH.

Taking into account the unknown location of the black hole, we
examined a field of $\sim\pm0.5''$ in right ascension and declination
around the center of IRS~13E with a grid step of $0.01''$.  The
residual proper motion of each star results in a required lower mass
for the IMBH -- dependent on its location -- in order to bind the star
to IRS~13E. Thus, one can obtain four maps for the sources E1, E2, E3,
and E4.  The four maps were combined by taking the maximum (otherwise
the system would not be bound) of the calculated mass limits at a
given location. We show the resulting map of lower mass limits in the
right panel of Fig.~\ref{Fig:Masses}. The significance of the
calculated masses over the field is $\geq4$\,$\sigma$.  Since E1 has
the proper motion with the strongest deviation from the mean proper
motion of the ensemble, it is the star that gives the strongest
constraints. There is a narrow minimum in the mass map for black hole
locations close to E1 ($\sim$0.1$''$ NW of E1). In this most
conservative case, the mass of the black hole would have to be larger
than $7000$\,\solm~(with a significance of 4\,$\sigma$). If the black
hole were located between E2 and E4, the location used by
\citet{Maillard2004A&A} for their mass estimate, it would have to have
a mass $>15000$\,\solm~(significance $>10\sigma$).

These are only lower limits that rely on projected velocities and
conservative estimates. If IRS~13E is indeed a bound system, a more
realistic mass might be derived with the LM estimator. Again, there is a
minimum near E1 for the black hole mass. However, even in this
conservative case, the LM estimator gives a mass of
$\sim$50000$\pm15000$\,\solm.

If we include the source E6 in the above described analysis, the
required black hole masses would at least double, due to the increased
velocity gradient across the complex. Hence, we conclude that it is
safe to assume that the required mass to bind the members
of the IRS~13E complex must be at least
$10^{4}$\,\solm. Assuming the extreme case that the 4 brightest stars
in IRS~13E are similarly massive as the binary IRS~16SW
\citep[$\ge$100\,\solm, see][]{Ott1999}, it appears conservative to assume
that the stellar mass in IRS~13E does not exceed $10^{3}$\,\solm~and
can thus be neglected in this analysis.

\section{Discussion}

IRS~13E is located at a projected distance of $\sim$$3.5''$ or
$\sim$130\,mpc from the $~3.5\times10^{6}$\,\solm~ black hole Sgr~A*.
It may be associated with the counter-rotating disk of young stars
\citep{Genzel2003ApJ}. In this case, it should be located at
$\sim$120\,mpc behind the plane of the sky. With a radius of
$\sim$$0.25''$, a mass of roughly 1000\,\solm~ then would be
sufficient to protect the system from tidal disruption. However, as
the analysis above shows, the real constraints for binding IRS~13E
gravitationally result from the intrinsic proper motions of the
sources in this complex.  We took into account the unknown location of
the hypothetical black hole and various possibilities for the systemic
motion of IRS~13E in the gravitational potential of Sgr~A*. In the
most conservative case, the minimum mass to bind IRS~13E was found to
be $7000\pm1800$\,\solm~(with a significance of 4\,$\sigma$). This
would confine the IMBH to a narrow region $\sim0.1''$ NW of
E1. Outside this region the required mass is greater than
$10^4$\,\solm. The LM mass estimator results in even higher masses.

It is generally accepted that accreting black holes are associated
with non-thermal radio and X-ray emission. One may argue that an IMBH
in IRS~13E could be ``starved'', similar to Sgr~A* \citep[see,
e.g.,][]{Melia2001ARA&A}, with its emission below the detection limit
of current telescopes. However, there are several pieces of evidence
that the IRS~13E cluster is closely associated with the gas and dust
of the mini-spiral: First, there is the close spatial relation between
IRS~13E and emission from warm dust \citep[see, e.g., the
high-resolution AO 3.8\,$\mu$m images
in][]{Eckart2004ApJ}. \citet{Paumard2004A&A} present an analysis of
the mini-spiral that also supports a close relation between IRS~13E
and the ISM. Also, the stars in IRS~13E, with
some of them being WR stars, may present favorable sources of gas and
dust.  \citet{Moultaka2004A&A} and \citet{Moultaka2005} present
spectroscopic evidence for absorption/emission due to gas and dust
intrinsic to IRS~13E and for interaction between winds from IRS~13E
and the surrounding ISM.  Hence, the situation for a hypothetical
black hole in IRS~13E appears to be different from Sgr~A*, with plenty
of material available for accretion.

\begin{figure}
\center
\includegraphics[angle=0,width=0.8\textwidth]{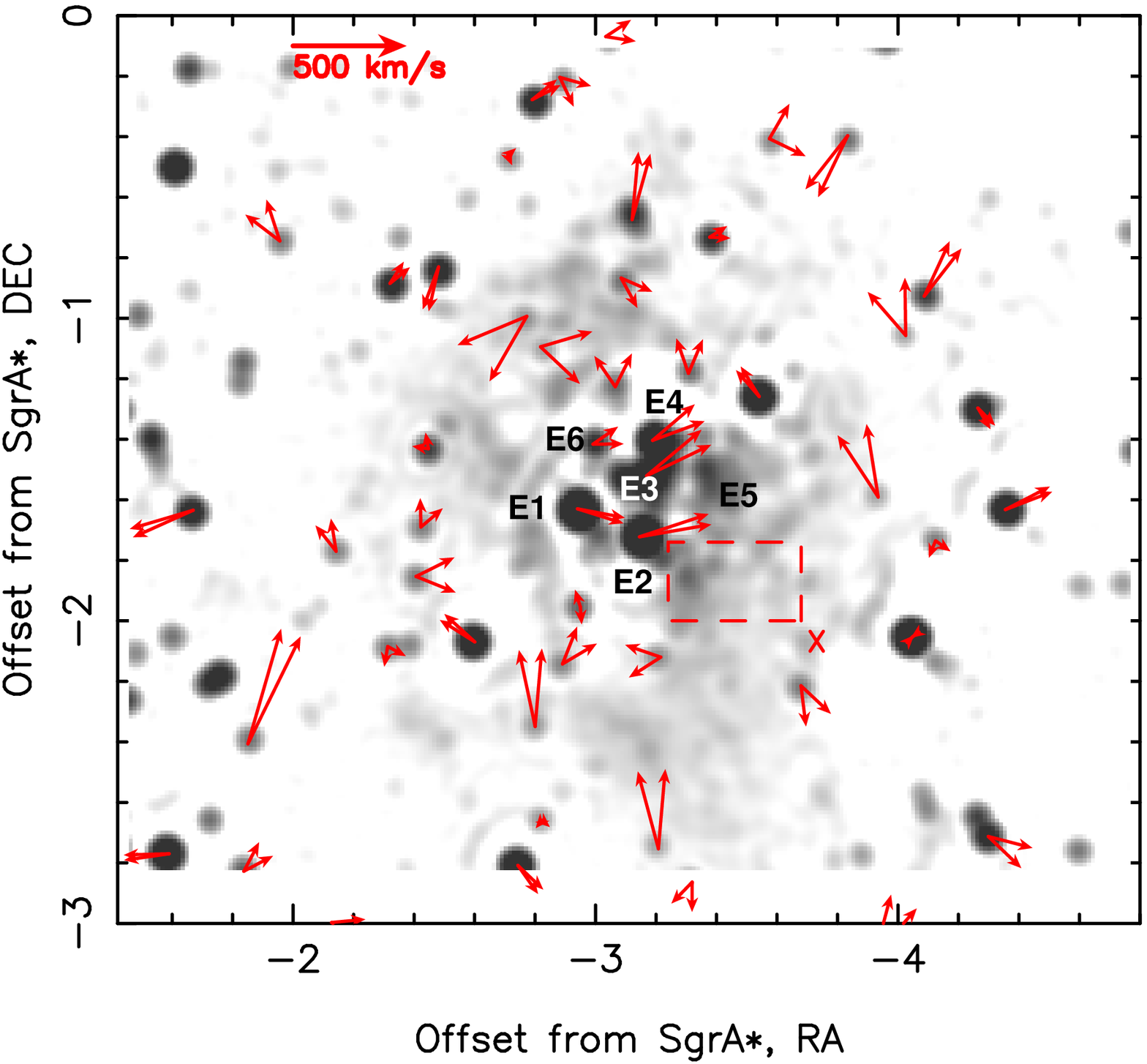}
\caption{Proper motions of stars in the vicinity of IRS~13E. 
 The underlying image is a LR-deconvolved, beam-restored (60\,mas FWHM
  Gaussian beam) NACO K-band image from 8 July 2004. Two arrows are
  shown for each source, indicating the $\pm$3~$\sigma$ uncertainty of
  the direction of its proper motion. The lengths of the arrows
  correspond to the magnitude of the respective velocity. The dashed
  box marked with an ``X'' designates the position (1\,$\sigma$ uncertainty) of the X-ray source
  near IRS~13E reported by \citet{Baganoff2003ApJ}.
  \label{Fig:Propmot}}
\end{figure}

\begin{figure}
\center
\includegraphics[angle=0,width=0.8\textwidth]{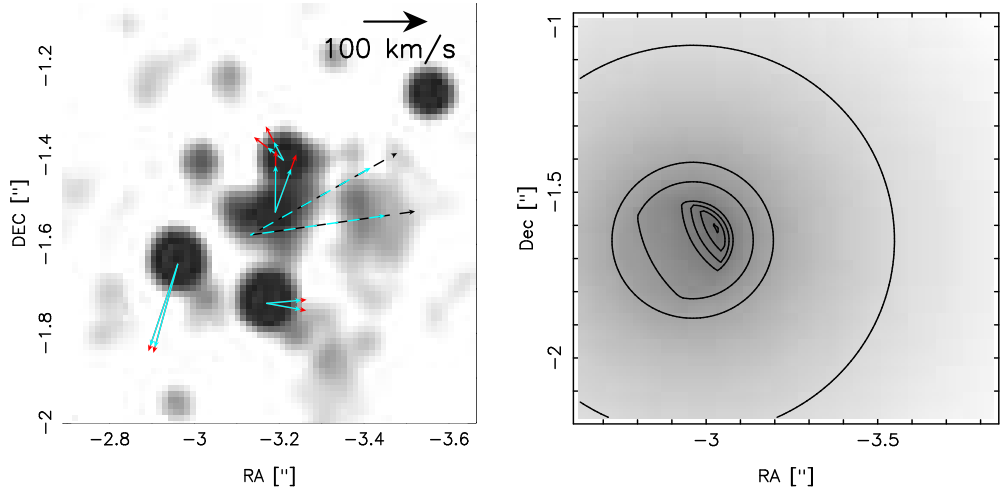}
\caption{Left panel: Zoom into the image shown in
  Fig.~\ref{Fig:Propmot}. Shown are the residual proper motions of the
  IRS~13E sources after subtraction of the average projected velocity
  (indicated by the dashed arrows) of the sources E1, E2, E3, and
  E4. The opening angles define the 1\,$\sigma$ uncertainty of the
  directions. The superposed light grey arrows give the magnitude of
  the velocities after subtraction of a 1\,$\sigma$ error. Right
  panel: Minimum mass required for an IMBH in order to gravitationally
  bind IRS~13E. This mass depends on the location of the IMBH and was
  calculated with the residual velocities of the IRS~13E
  sources. Contours at 7, 8, 9, 10, 15, 20, and
  50$\times10^{3}$\,\solm, growing larger toward lighter shades of
  gray. The significance of the masses is 4-16\,$\sigma$.
  \label{Fig:Masses}}
\end{figure}

As for mm/radio emission, \citet{Eckart2004ApJ} show that the 13\,mm
point source observed by \citet{ZhaoGoss1998ApJ} in the IRS~13E
cluster is most likely due to thermal emission from the 4$\mu$m-excess
source E3c in the center of IRS~13E, a dusty WR star \citep[see
also][]{Moultaka2005}.  The two components E3A and E3B are also
proposed as dusty WR stars in the analysis by
\citet{Maillard2004A&A}. As for X-ray emission,
\citet{Baganoff2003ApJ} report an X-ray source that they associate
with IRS~13E. Using the position of this source and of Sgr~A* as
listed in their Table~1, we located this X-ray source in the NIR
reference frame. Its position is marked with a 1\,$\sigma$ error box
in Fig.~\ref{Fig:Propmot}. The X-ray source appears offset from
IRS~13E as also discussed by \citet{Maillard2004A&A}
. \citet{Coker2002A&A} interpret the spectrum of the source as
consistent with that of a post-LBV WR colliding wind binary (although
none of the bright IRS~13E members appears to be associated with the
source, see Fig.~\ref{Fig:Propmot}). As for a very recent analysis of
this X-ray source, it appears to be located a few $0.1''$ further
north than indicated in Fig.~\ref{Fig:Propmot}, so that E5 lies within
the positional uncertainty. No variability of the source was found and
it appears to be explained well by colliding winds (F. Baganoff,
priv. comm.).  Therefore there appears to be no compelling evidence
for an accreting IMBH in IRS~13E. Furthermore,
spectroscopy at 4\,$\mu$m \citep{Moultaka2004A&A} and 5\,$\mu$m
\citep{Moultaka2005} does not reveal any strong broad lines or
emission from highly excited ions as they might be expected near an
accreting black hole.

Infall and dissolution of a massive cluster into the central parsec
seems an attractive hypothesis for delivering young stars to
the region near Sgr~A*. Generally, the conditions required for a
cluster to form near the GC and to fall in towards the central parsec
to deposit young, massive stars there within their lifetime, are
extreme, i.e., high core densities and large cluster masses \citep[see
discussions
in][]{Gerhard2001ApJL,Genzel2003ApJ,Kim2003ApJ,Kim2004ApJL,Maillard2004A&A}. An
IMBH in the core of such a cluster was suggested by
\citet{Hansen2003ApJ} as a means of increasing the efficiency of
dynamical friction on the cluster, and thus to relax somewhat the
demands on its total mass and core density. \citet{Kim2004ApJL},
however, conclude that an IMBH in the cluster core must comprise an
unrealistic $\sim$10\% of the total cluster mass in order for this
mechanism to be effective.

From the proper motions presented here, the required minimum mass for
a hypothetical IMBH in order to bind the members of IRS~13E was
derived to be $\gtrsim10^{4}$\,\solm. \citet{Portegies2002ApJ} analyze
the growth of an IMBH by runaway collisions in a dense stellar cluster
and conclude that such an object may form in dense clusters and
contain of the order $0.1\%$ of the total stellar mass. Given the high
mass estimate for an IMBH in IRS~13E derived in this work, this would
mean an unrealistically high mass of $>10^{6}$\,\solm~ for the
progenitor cluster. In this context it is also important to point out
the results of \citet{Reid2004ApJ} that exclude the existence of a
secondary black hole with masses $\gtrsim10^{4}$\,\solm~ and
semi-major axes between $10^{3}$ and $10^5$\,AU (corresponding to
$\sim$0.1$''$ to 12.5$''$ angular distance from Sgr~A*). We cannot yet
exclude an IMBH of up to a few thousand solar masses in IRS~13E, but
our results suggest that in this case IRS~13E must be in the process
of dissolution.  In this context, we would like to point out the
recent work by Levin, Wu and Thommes (astro-ph/0502143). They have
simulated the infall and dissolution of a stellar cluster with an IMBH
in the GC and concluded that a hypothetical IMBH that binds the
IRS~13E complex could not have delivered all of the young stars in the
GC to their present location.

The positions and proper motions of the IRS~13E stars are consistent
with their being part of the counter-clockwise disk/ring of young
stars \citep[][see also T.~Paumard et al., 2005, in
preparation]{Genzel2003ApJ}. As for the unusual clustering of the
stars in IRS~13E, the complex may be, on the one hand, a
(quasi-)permanent feature, i.e. gravitationally bound by a
hypothetical IMBH. Our analysis shows that in this case, if the star
E1 is indeed part of the complex, the IMBH would have to be unsually
massive. This makes its existence questionable because of the required
large mass of the progenitor cluster and because of the lack of a
clear identification of a non-thermal, variable X-ray source centered
within IRS~13E. However, we cannot completely discard the IMBH
hypothesis. On the other hand, IRS~13E may be a temporary feature,
i.e., either a chance association (which is appears somewhat unlikely,
even if the stars are part of a disk/ring) or a cluster in the process
of dissolution.  At this point, we would like to point out that an
unusual concentration of co-moving stars has recently also been
identified among the members of the IRS~16 complex in the
clockwise rotating disk of stars \citep[see][]{Lu2004AAS}. Detailed
future studies of the proper motions, line-of-sight velocities,
distribution, and spectral types of the stars in the central parsec of
the GC nuclear cluster are clearly needed in order to understand
better the phase-space clustering of the stars in the IRS~13E and
IRS~16 complexes.

\acknowledgments \emph{Based on observations obtained at the
Gemini Observatory, which is operated by the Association of
Universities for Research in Astronomy, Inc., under a cooperative
agreement with the NSF on behalf of the Gemini partnership: the
National Science Foundation (United States), the Particle Physics and
Astronomy Research Council (United Kingdom), the National Research
Council (Canada), CONICYT (Chile), the Australian Research Council
(Australia), CNPq (Brazil) and CONICET (Argentina).}

\clearpage


\clearpage

\begin{table}
\begin{center}
\caption{Positions and velocities of sources in IRS~13E (labeling see
  Fig.~\ref{Fig:Propmot}).Positions are given as offsets from Sgr~A*
  in arc-seconds at the epoch 2004.73. \label{Tab:Vels}}
\begin{tabular}{ccccc}
Source ID & RA [$''$] & Dec [$''$] & v$_{RA}$ [km\,s$^-1$] & v$_{Dec}$ [km\,s$^-1$]\\
\tableline\tableline
E1 & -2.961 & -1.645 & -206$\pm$4  & -51$\pm$4\\
E2 & -3.171 & -1.732 & -311$\pm$4  & 77$\pm$8\\
E3 & -3.191 & -1.527 & -265$\pm$17 & 171$\pm$11\\
E4 & -3.210 & -1.411 & -210$\pm$9  & 124$\pm$16 \\
E6 & -3.013 & -1.422 &-128$\pm$16 & 41 $\pm$13 \\
\tableline
\end{tabular}
\end{center}
\end{table}


\end{document}